%% file: main.tex
\documentclass[epjCONF,columns]{svjour}
\usepackage{graphics}
\usepackage[varg]{txfonts} 
\usepackage[latin1]{inputenc}
\usepackage{url}

\usepackage{graphicx}  
\usepackage{xspace}
\usepackage{color}
\usepackage{colortbl}
\usepackage{amssymb}
\usepackage{amsfonts}
\usepackage{amstext}
\usepackage{upgreek}
\usepackage{ifthen} 
\newboolean{uprightparticles}
\setboolean{uprightparticles}{false} 
\newboolean{pdflatex}
\setboolean{pdflatex}{false} 
\newcommand{\avg}[1]{\left< #1 \right>} 
\usepackage{lhcb-symbols-def}
\input{os_definitions.tex}

\input{sin2beta_definitions.tex}

\session-title{2011 Hadron Collider Physics symposium (HCP-2011)}
\begin{document}
\title{Search for CP violation in $B^0 \rightarrow J/\psi K^0_S$  decays with first LHCb data}
\author{Murilo Rangel \thanks{\email{rangel@if.ufrj.br}} \\
on behalf of LHCb collaboration}
\institute{Universidade Federal do Rio de Janeiro}
\abstract{
We report a measurement of the CP violation in $B^0 \rightarrow J/\psi K^0_S$ decays.
We perform a time-dependent analysis of the decays reconstructed in $35 pb^{-1}$ of LHCb data that was taken in 2010. 
We measure the CP asymmetry parameter 
$S_{J/\psi K_S^0} = 0.53^{+0.28}_{-0.29}$ (stat) $\pm 0.05$ (syst), 
which is connected to the CKM angle $\beta$.
} 
\maketitle
\section{Introduction}
\label{intro}

The decay $B^0 \rightarrow J/\psi K^0_S$ is well known as the gold-plated mode for the study of CP violation 
in the $B^0$ meson decays to a CP eigenstate common to both $B^0$ and $\bar{B}^0$, allowing for interference 
through oscillation. Therefore, measurements of the decay $B^0$ and $\bar{B}^0$ have a good sensitivity to 
the CKM angle $sin(2\beta)$, which is connected to the parameters \S and \C:

\begin{eqnarray}
{\cal A}_{\jpsi\KS}(t)  \equiv
  \frac{\Gamma(\Bzb(t)\to\jpsi\KS) - \Gamma(\Bz(t)\to\jpsi\KS)}
       {\Gamma(\Bzb(t)\to\jpsi\KS) + \Gamma(\Bz(t)\to\jpsi\KS)} 
\nonumber
\\
 = \S\sin(\dmd t) - \C\cos(\dmd t) ~,
\end{eqnarray}

where $\dmd = m_{B_H^0} - m_{B_L^0}$ is the mass difference between the mass eigenstates.
We have neglected the decay width difference between the
mass eigenstates of the \Bz system.
The connection of the \S and \C parameters to the CKM angle is

\begin{equation}
\S \simeq \sqrt{1-{\C}^2}\sintwobeta ~.
\end{equation}

In the Standard Model, both \CP violation in mixing and direct
\CP violation are negligible in $b\to\ccbar s$ decays. As a consequence
the cosine term vanishes, implying

\begin{equation}
\SJpsiKS \simeq \sintwobeta ~.
\end{equation}

During the last decade the B-factories BABAR and Belle reached outstanding 
precision in the measurement of $S_{J/\psi K_S^0}$. The most recent BABAR measurement
reports $S_{J/\psi K_S^0} = 0.663 \pm 0.039$~(stat)~$\pm~0.012$~(syst) \cite{babar}.
The most recent Belle result is $S_{J/\psi K_S^0} = 0.642 \pm 0.031$~(stat)~$\pm 0.017$~(syst) \cite{belle}.
Currently, the world average \cite{hfag} is $sin 2\beta = 0.673 \pm 0.023$. 

The measurement presented in this note \cite{ConfNote} with the first LHCb data is an important step 
to demonstrate the potential of the LHCb experiment in this topic and it demonstrates 
that the flavour tagging algorithms are under control.
We measure \SJpsiKS under the assumption that $\C=0$,
and we quote the resulting value of \C if the Standard Model
constraint is relaxed.

\section{Analysis Strategy}
\label{sec:strat}

The results presented are based on data collected 
with the LHCb detector at the LHC collider at a center-of-mass
energy of $\sqrt{s}=7\tev$.
Details on the LHCb experiment can be found in elsewhere ~\cite{LHCb}.
The data set analyzed has an integrated luminosity of $\approx 35\invpb$.
We use Monte Carlo (MC) simulated samples that are based on the \textsc{Pythia} 6.4
generator \cite{Pythia}.
The \textsc{EvtGen} package \cite{EvtGen} was used to generate hadron decays
and the \textsc{Geant4} package \cite{Geant4} for detector simulation. 

We reconstruct \jpsi candidates in the decay mode $\jpsi\to\mu^+\mu^-$,
from pairs of opposite sign tracks that have a
transverse momentum of $\pt > 500\mevc$ each and 
particle identification signatures consistent with those of
muons. The invariant mass of the pair must be compatible with the known
\jpsi mass.
\KS candidates are reconstructed through their decay into $\pip\pim$
with take pairs of oppositely charged tracks.
An additional requirement of $L / \sigma_L > 5$ is made on the
\KS candidate, where $L$ is the distance between the \KS decay vertex and the $B$
decay vertex and $\sigma_L$ is the uncertainty of $L$.
We constrain the invariant masses of the reconstructed \jpsi and \KS
candidates to their known masses.

The initial $B$ flavour is determined by the combination of various
tagging algorithms. These either determine the flavour of the non-signal
$b$ hadron produced in the event (\emph{opposite side}, OS), or they
search for an additional pion accompanying the signal \Bz or \Bzb
(\emph{same side}, SS$\pi$). There are four tagging algorithms that 
use the charge of the lepton ($\mu$, $e$) from
semileptonic $B$ decays, or that of the kaon from the $b\to c\to s$
decay chain, or the charge of the inclusive secondary vertex
reconstructed from $b$ decay products. All of these algorithms have
an intrinsic mistag rate, due to picking up tracks from the underlying
event, or due to flavour oscillations of neutral tag $B$ mesons.
For each signal \Bz candidate the tagging algorithms also predict
the mistag probability \mistag. For this various kinematic
variables such as momenta and angles of the tagging particles 
are combined into neural networks. The neural networks are trained on MC simulated events.

The flavour asymmetry that is accessible in \BdJKS decays directly
depends on the dilution $D$ due to the mistag probability for each signal candidate, 
$\avg{D^2}=\frac{1}{N} \sum_i (1-2\mistag_i)^2$.
Its statistical precision is proportional to the inverse square root of 
the effective tagging efficiency $\varepsilon_{\rm eff}$,
\begin{equation}
   \varepsilon_{\rm eff} = \etag \avg{D^2}\, ~,
\end{equation}
where \etag\ is the probability that a tagging decision is found.
The tagging algorithms are optimized for highest $\varepsilon_{\rm eff}$ on data, 
using the self-tagging decays $\Bp\to\jpsi\Kp$ and $\Bd\to\Dstarm\mup\nu$. 
The estimated mistag probability is calibrated on these same channels.
The effective tagging efficiency is measured to be $\varepsilon_{\rm eff}=(2.82 \pm 0.87)\%$.

To extract the  $S_{J/\psi K_S^0}$ parameter, we perform a simultaneous, extended unbinned maximum 
likelihood fit to the proper time and the invariant mass distributions:

\begin{equation}
   \mathcal{L}(\vec{\lambda}) = \frac{e^{-N} N^n}{n!} \prod_{s} 
   \prod_{i=1}^{N^s} \mathcal{P}^s(\vec{x}_i; \vec{\lambda}_s) \ . 
\end{equation}

We minimize $-\ln\mathcal{L}$ to find optimal values for the
fit parameters $\vec{\lambda}$.

The fit is simultaneous in four subsamples $s$. These subsamples are defined
by whether the candidates were triggered by a lifetime unbiased (``U'') 
or a lifetime biased (``B'') decision; and by whether or not a tagging
decision is available (``u'' for untagged, ``t'' for tagged).
Each subsample contains $N^s$ events, and $n = \sum_s N^s$.

We consider four observables:
the reconstructed mass $m$ of the \Bd candidate 
($5.15\gevcc<m<5.4\gevcc$), its proper time $t$ ($-1\ps<t<4\ps$),
the flavour tag decision $d$, and the combined per-event mistag prediction \mistag.
The flavour tag $d$ can take the discrete values $d=1$ if 
tagged as initial \Bd and $d=-1$ if tagged as initial \Bdb.

The probability density functions (p.d.f.s) $\mathcal{P}^s$ consist
of three components, signal (S), prompt background (P), and long lived background (L).
The mass p.d.f. of the signal component consists of a single Gaussian.
We assume both background components have similar mass distributions, and we use
the same parameterization for both. It is modeled as an exponential with
a single shape parameter. 
The proper time p.d.f. of the signal component can be written as
$\mathcal{P}_S(t,d,\mistag) = \mathcal{P}_S(t,d|\mistag) \cdot \mathcal{P}_S(\mistag)$.
The first term is a conditional p.d.f. as it depends on the value of \mistag, 
the second term describes the distribution of \mistag. The background 
parameterization of the proper time factorizes,
$\mathcal{P}_B(t,d,\mistag) = \mathcal{P}_B(t,d) \cdot \mathcal{P}_B(\mistag)$,
where $B = P, L$.


In the fit to the \BdJKS channel we fix the mixing frequency \dmd\ to its nominal value of $\dmd = (0.507
\pm 0.005) \cdot 10^{12} \, \hbar s^{-1}$~\cite{PDG}, and $\C=0$. In total, there are
27 floating parameters: the \CP parameter \S, the \Bd lifetime $\tau$, the \Bd mass $m_0^S$, twelve
event yields, four parameters of the long-lived proper time background, five parameters
of the time resolution, the mass signal resolution $\sigma_{m}^S$, and two parameters of the mass background shape.
We have checked the fit implementation on a large sample of MC generated signal
events, where we find good agreement with the generated values.

The measurement was performed using a ``blind'' analysis technique to minimize unconscious experimenter bias, 
the parameter of interest was encrypted in the likelihood fit. 
Only after the full analysis strategy was developed and proved to be stable, 
the encryption was removed, unblinding the true result.

\subsection{Results}
\label{sec:results}

\begin{figure}[!htb]
\center
\includegraphics[width=0.475\textwidth]{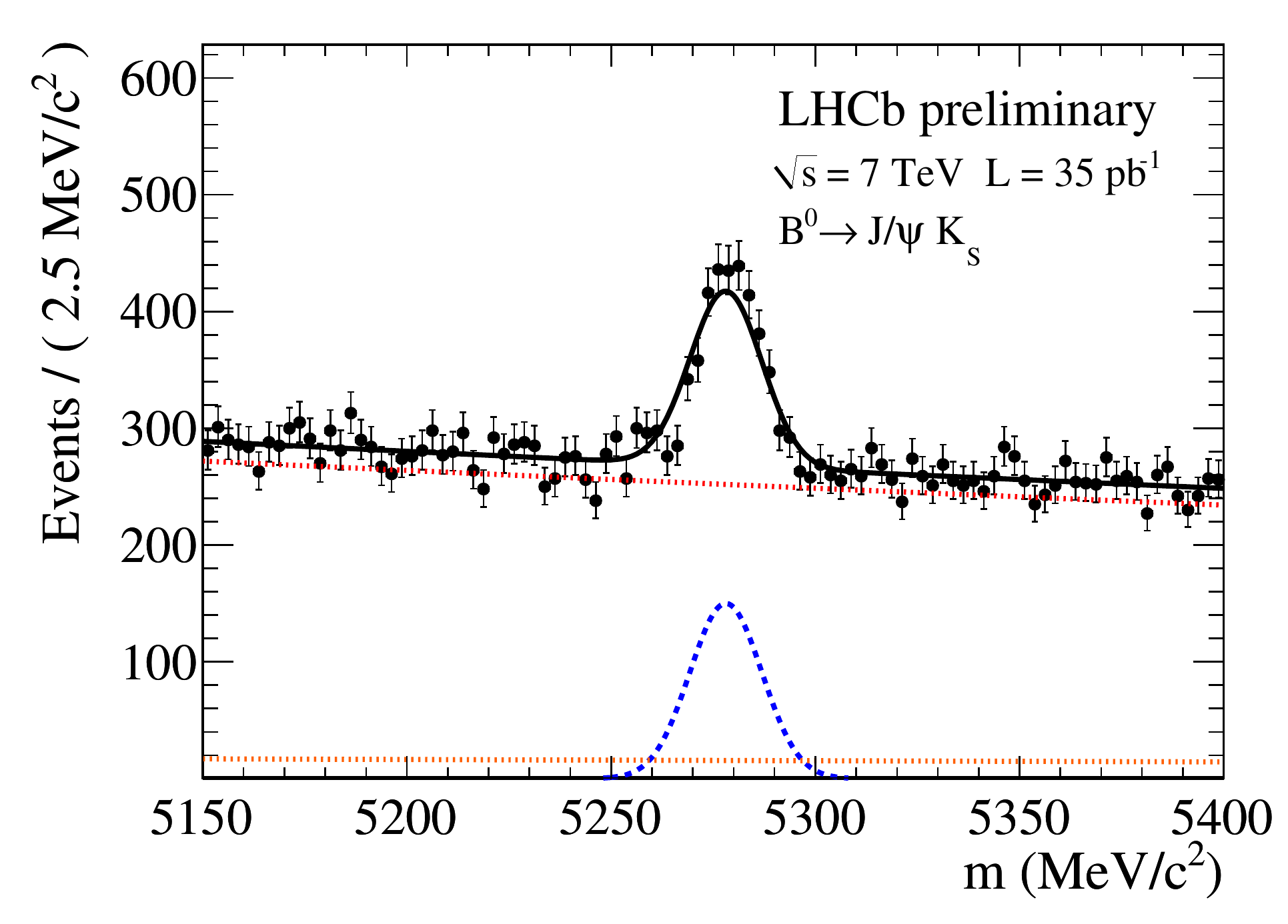}
\includegraphics[width=0.475\textwidth]{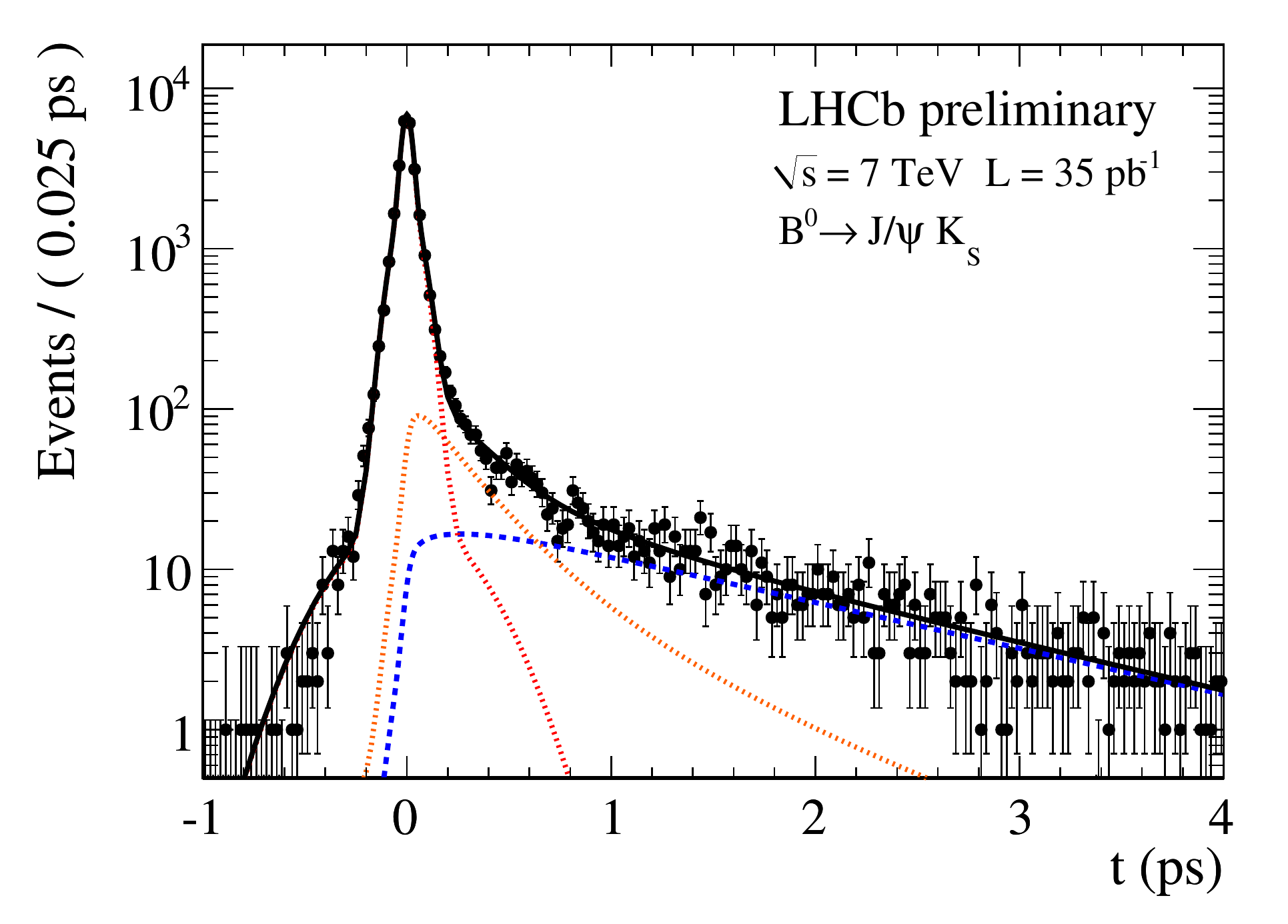}
\caption{Reconstructed mass (left) and proper time (right) distributions
of \BdJKS candidates. Overlayed are projections of the component p.d.f.s
used in the fit:
full p.d.f. (solid black),
signal (dashed blue),
prompt background (dash-dotted red),
long lived background (dotted orange).}
\label{fig:fitresult}
\end{figure}

\begin{figure}[!htb]
\center
\includegraphics[width=0.475\textwidth]{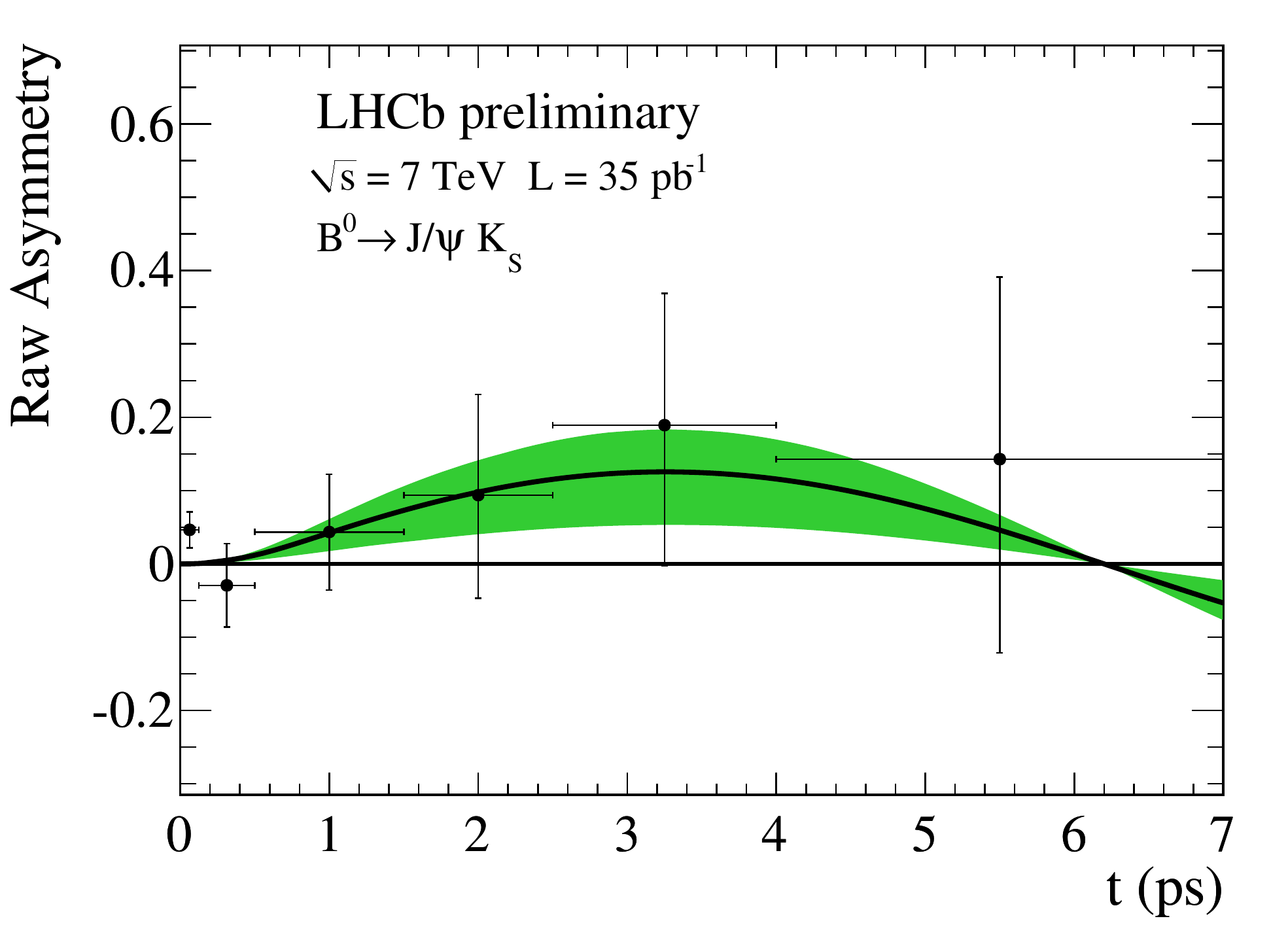}
\caption{Time dependent raw \CP asymmetry in \BdJKS. The solid curve is the full p.d.f. (signal and background)
overlayed onto the data points. The green band corresponds to the one standard deviation 
statistical error.}
\label{fig:asymmetry}
\end{figure}

The result of the maximum likelihood fit to the full data sample
is summarized in Tables~\ref{tab:fitresults} and~\ref{tab:fitresultsyields}.
The mass and proper time distributions and the fit
projections are shown in Figure~\ref{fig:fitresult}. Figure~\ref{fig:asymmetry}
shows the resulting time dependent raw asymmetry, which contains all fit components.
The asymmetry in the lowest proper time bins is therefore dominated by the backgrounds, whereas the measured
asymmetry in the high proper time
bins is dominated by signal events.
The measured value of \S is
\begin{equation}
	\S = 0.53^{+0.28}_{-0.29} \ ,
\end{equation}
where the error
is statistical only. We find the global correlation coefficient of \S to be $\rho(\S)=0.016$.
We also perform the nominal fit without the Standard Model constraint $\C=0$.
In this case, we find
\begin{equation}
	\C = 0.28 \pm 0.32 \ , \\
	\S = 0.38 \pm 0.35 \ ,
\end{equation}
again quoting statistical errors only.
The correlation coefficient between the parameters is $\rho(\S,\C)=0.53$.
Their correlations to other parameters are negligible.

\begin{table}
\caption{Fit result of the nominal fit to the full \BdJKS data sample.}
  \label{tab:fitresults}
  \centering
  \begin{tabular}{lcc}
  \hline  \hline Parameter& Unit              & Fitted Value\\ \hline
  \\ [-3.5mm]
  \parProptimeSf            &                 & $  0.53_{-0.29}^{+0.28}$               \\
  \parSigMassMean           & \mevcc          & $  5278.13 \pm 0.29$            \\
  \parSigMassSigma          & \mevcc          & $  8.82 \pm 0.24$               \\ 
  \parSigProptimeTau        & \ps             & $  1.517 \pm 0.046$            \\ [1.5mm] \hline \\ [-3.5mm]
  \parBkgMassExpoBi         & $(\mevcc)^{-1}$ & $ -8.71 \pm 3.8 \cdot 10^{-4}$   \\
  \parBkgMassExpoUb         & $(\mevcc)^{-1}$ & $ -5.86 \pm 0.87 \cdot 10^{-4}$\\
  \parLbgProptimeDecFracU   &                 & $  0.836 \pm 0.054$            \\
  \parLbgProptimeTauOneU    & \ps             & $  0.221 \pm 0.036$            \\
  \parLbgProptimeTauTwoU    & \ps             & $  1.04 \pm 0.24$               \\
  \parLbgProptimeTauB       & \ps             & $  0.482 \pm 0.029$            \\
  \parrProptimeGMFracOne    &                 & $  0.500 \pm 0.019$            \\
  \parrProptimeGMFracTwo    &                 & $  0.477 \pm 0.017$            \\
  \parrProptimeSigmaOne     & \ps             & $  0.02522 \pm 0.00066$      \\
  \parrProptimeSigmaTwo     & \ps             & $  0.0685 \pm 0.0016$         \\
  \parrProptimeSigmaThree   & \ps             & $  0.293 \pm 0.019$            \\ [1.5mm] \hline\hline
\end{tabular}
\end{table}

\begin{table}[!htb]
  \caption{Fitted event yields in the full \BdJKS data sample.}
  \label{tab:fitresultsyields}
  \centering
  \begin{tabular}{clcc}
  \hline  \hline Sample & Parameter & Fitted Value  \\ \hline
        & \parSigYieldUT           & $  221 \pm 17$     \\
   U,t  & \parPbgYieldUT           & $  3218 \pm 62$    \\
        & \parLbgYieldUT           & $  309 \pm 33$     \\ \hline
        & \parSigYieldBT           & $  59.8 \pm 8.7$  \\  
   B,t  & \parPbgYieldBT           & $  164 \pm 14$     \\
        & \parLbgYieldBT           & $  102 \pm 12$     \\ \hline  
        & \parSigYieldUU           & $  767 \pm 32$     \\
   U,u  & \parPbgYieldUU           & $  21134 \pm 161$ \\  
        & \parLbgYieldUU           & $  807 \pm 79$     \\ \hline
        & \parSigYieldBU           & $  279 \pm 18$     \\
   B,u  & \parPbgYieldBU           & $  747 \pm 30$     \\
        & \parLbgYieldBU           & $  339 \pm 23$     \\ \hline \hline
\end{tabular}
\end{table}

\section{Conclusion}
\label{sec:conclusion}

The final result on the \CP violation parameter \S is

\begin{equation}
	\S = 0.53^{+0.28}_{-0.29}\stat \pm 0.05\syst \ .
\end{equation}

This result is compatible with the World Average, and 
dominated by the statistical uncertainty. We calculate the statistical
significance of a non-zero \CP violation from the likelihood
ratio of a test fit, in which we fix $\S=0$, to be

\begin{equation}
	S = \sqrt{2 \, \ln(\mathcal{L}_{\rm nom} / \mathcal{L}_{\rm null}) \,} = 1.8 ~.
\end{equation}

This is the first \CP violation result in the golden channel \BdJKS in LHCb.

\end{document}

%% file: os_definitions.tex
\newcommand{\comment}[1]{\par\noindent {\em\small [#1]}}
\renewcommand{\comment}[1]{}

\newcommand{\particle}[1]{{\ensuremath{\rm #1}}}

\renewcommand{\b}{\particle{b}}

\newcommand{\BdJKS}{\decay{\Bd}{\jpsi\KS}}               



%% file: sin2beta_definitions.tex
\newcommand{\sintwobeta}{\ensuremath{\sin 2\beta}\xspace}

\newcommand{\CJpsiKS}{\ensuremath{C_{\jpsi\KS}}\xspace}
\newcommand{\SJpsiKS}{\ensuremath{S_{\jpsi\KS}}\xspace}

\renewcommand{\S}{\SJpsiKS}
\renewcommand{\C}{\CJpsiKS}

\newcommand{\parProptimeSf}[0]{\ensuremath{\SJpsiKS}}

\newcommand{\parSigMassMean}[0]{\ensuremath{m_\text{S}}}
\newcommand{\parSigMassSigma}[0]{\ensuremath{\sigma_{\text{S},m}}}
\newcommand{\parSigProptimeTau}[0]{\ensuremath{\tau}}
\newcommand{\parBkgMassExpoBi}[0]{\ensuremath{\alpha^\text{B}_{m}}}
\newcommand{\parBkgMassExpoUb}[0]{\ensuremath{\alpha^\text{U}_{m}}}
\newcommand{\parLbgProptimeDecFracU}[0]{\ensuremath{f^\text{U}_{\text{L},t}}}
\newcommand{\parLbgProptimeTauOneU}[0]{\ensuremath{\tau^{\text{U}}_{\text{L},1}}}
\newcommand{\parLbgProptimeTauTwoU}[0]{\ensuremath{\tau^{\text{U}}_{\text{L},2}}}
\newcommand{\parLbgProptimeTauB}[0]{\ensuremath{\tau^{\text{B}}_{\text{L}}}}
\newcommand{\parLbgYieldBT}[0]{\ensuremath{N^\text{B,t}_\text{L}}}
\newcommand{\parLbgYieldBU}[0]{\ensuremath{N^\text{B,u}_\text{L}}}
\newcommand{\parLbgYieldUT}[0]{\ensuremath{N^\text{U,t}_\text{L}}}
\newcommand{\parLbgYieldUU}[0]{\ensuremath{N^\text{U,u}_\text{L}}}  
\newcommand{\parPbgYieldBT}[0]{\ensuremath{N^\text{B,t}_\text{P}}}  
\newcommand{\parPbgYieldBU}[0]{\ensuremath{N^\text{B,u}_\text{P}}}
\newcommand{\parPbgYieldUT}[0]{\ensuremath{N^\text{U,t}_\text{P}}}
\newcommand{\parPbgYieldUU}[0]{\ensuremath{N^\text{U,u}_\text{P}}}
\newcommand{\parSigYieldBT}[0]{\ensuremath{N^\text{B,t}_\text{S}}}
\newcommand{\parSigYieldBU}[0]{\ensuremath{N^\text{B,u}_\text{S}}}
\newcommand{\parSigYieldUT}[0]{\ensuremath{N^\text{U,t}_\text{S}}}
\newcommand{\parSigYieldUU}[0]{\ensuremath{N^\text{U,u}_\text{S}}}

\newcommand{\parrProptimeGMFracOne}[0]{\ensuremath{f_{\mathcal{R},1}}}
\newcommand{\parrProptimeGMFracTwo}[0]{\ensuremath{f_{\mathcal{R},2}}}

\newcommand{\parrProptimeSigmaOne}[0]{\ensuremath{\sigma_{\mathcal{R},1}}}
\newcommand{\parrProptimeSigmaTwo}[0]{\ensuremath{\sigma_{\mathcal{R},2}}}
\newcommand{\parrProptimeSigmaThree}[0]{\ensuremath{\sigma_{\mathcal{R},3}}}